# Optimization of phase retrieval in the Fresnel domain by the modified Gerchberg-Saxton algorithm

SOHEIL MEHRABKHANI,[1,2,*] AND MELVIN KÜSTER,[1]

[1]Terahertz-Photonics group, Institut für Hochfrequenztechnik, TU Braunschweig, Schleinitzstraße 22, 38106 Braunschweig
[2]Lena, Laboratory for Emerging Nanometrology
*Corresponding author: soheil.mehrabkhani@ihf.tu-bs.de

**The modified Gerchberg-Saxton algorithm (MGSA) is one of the standard methods for phase retrieval. In this work we apply the MGSA in the paraxial domain. For three given physical parameters – i.e. wavelength, propagation distance and pixel size – the computational width in the Fresnel-Transform is fixed. This width can be larger than the real dimension of the input or output images. Consequently, it can induce a padding around the real input and output without given amplitude (intensity) values. To solve this problem, we propose a very simple and efficient solution and compare it to other approaches. We demonstrate that the new modified GSA provides almost perfect results without losing the time efficiency of the simplest method.**

## 1. INTRODUCTION

The full field information can be extracted from the complex amplitude, i.e. intensity and phase of the field. The intensity (amplitude) can easily be measured by the use of a CCD camera, however the measurement of the phase distribution is much more complicated. But, the phase information is required in many areas of optics. The conventional method to measure the phase is interferometry [1-3]. However, this requires a sophisticated experimental setup. More recent approaches use phase retrieval procedures [4]. In these methods the phase information is extracted from the intensity information. In the paraxial domain one of these methods is based on the exploitation of the transport-of-intensity (TIE) equation [5]. Since the coupling between phase and intensity is described by the TIE, an extraction of the phase information becomes possible by two intensity measurements at distinct distances from the object. However, its fundamental disadvantage is that it depends on the axial intensity derivative in the paraxial direction. Thus, the distance between the intensity measurements must be sufficiently short. Another, in principle iterative, approach is the Gerchberg-Saxton algorithm (GSA) [6,7] and its modified forms (MGSA) [8-15]. Compared to the TIE, the MGSA is much more flexible concerning the propagation distance. However, the drawback of the MGSA is its convergence and stagnation problem, occurring in some cases. In many MGSAs a propagation algorithm is needed. An exact propagation algorithm would be given by the Rayleigh-Sommerfeld (RS) [16] method. But, the RS suffers from time inefficiencies and is thus impractical in MGSA. A fast propagation method is the angular spectrum (AS) [16]. From the analytical point of view the AS is equivalent to RS [17,18] however, for numerical calculations it is only valid for propagation distances shorter than a critical distance [19]. Another fast method is the Fresnel-Transform (FRT) [16], which is valid in the paraxial domain. Because of the importance of the paraxial wave propagation, here the application of the MGSA in the paraxial domain is investigated. Restrictions due to the FRT and the consequence for some proposed methods to overcome these restrictions are presented and a new very efficient approach with almost perfect results will be introduced and compared to other methods.

## 2. FRESNEL-TRANSFORM

The FRT can be seen as the Rayleigh-Sommerfeld diffraction method in the paraxial approximation [19].

$$u(\boldsymbol{r}_2,z) = \frac{e^{i2\pi\frac{z}{\lambda}}}{i\lambda z} e^{i\frac{\pi r_2^2}{z\lambda}} \mathcal{F}\left\{u(\boldsymbol{r}_1,0)\, e^{ik\frac{\pi r_1^2}{z\lambda}}\right\}_{f=\frac{r_2}{\lambda z}} \quad (1)$$

Equation (1) can be used to calculate the scalar complex field $u(\boldsymbol{r}_2,z)$ at a distance $z$ in an output plane from a given field $u(\boldsymbol{r}_1,0)$ in a parallel input plane. The vectors $\boldsymbol{r}_1$ and $\boldsymbol{r}_2$ are the position vectors in the input and the output plane, respectively and $\lambda$ is the wavelength of the coherent wave. The advantage of the FRT is the possibility to utilize the Discrete-Fourier-Transform (DFT), which enables a fast calculation. The inverse of the pixel size $\delta x_1$ in the input plane defines the frequency bandwidth $\Delta f_x = 1/\delta x_1 = 2f_{max}$ and the sampling frequency [20]. Considering the same number of sampling points $N$ in the spatial and frequency domain, for the computational domain in the input and the output plane it can be written that $P_1 = N\delta x_1$ and $P_2 = N\delta x_2$. Here $N$ is the number of the sampling points according to the DFT in the FRT and not the pixel number of the object and image. According to the FRT the relationship between the frequency and the position in the output plane is:

$$\boldsymbol{f} = \frac{\boldsymbol{r}_2}{\lambda z} \Rightarrow (f_x, f_y) = \frac{(x_2, y_2)}{\lambda z} \quad (2)$$

Consequently, the frequency bandwidth $\Delta f_x$ and the width $P_2$ are related via $\Delta f_x = 2x_{2_{max}}/(\lambda z) = P_2/(\lambda z)$. Consequently, the fundamental relationship between the widths of computational domains in the FRT is $P_1 P_2 = N\lambda z$.

After some mathematical manipulations, for the given parameters $P_1$, $P_2$, $N$ and $\lambda$ there is only one critical propagation distance $z_{FT}$ [21] which satisfies the Nyquist criterion for the forward and backward transforms. Both are necessary in the propagation dependent MGSA and will be called Fresnel sampling condition.

$$z_{FT} = \frac{P_1 P_2}{N\lambda} = N\frac{\delta x_1 \delta x_2}{\lambda} \quad (3)$$

For the same pixel size $\delta x$ and same sampling number $N$ in the input and output plane the width of the computational domain in both planes is equal: $P_1 = P_2 = P$. It can be easily shown that for a given $z$, $\lambda$ and $\delta x$ the sampling number $N$ in the FRT is fixed at $N = \lambda z/\delta x^2$. Introducing this into Eq. (3) results in $z_{FT} = z$, which means that the calculated $N$ and the corresponding $P$ satisfy the Nyquist criterion in the FRT for the forward and backward direction. However, the consequence is that the width of the Fresnel computational domain $P$ can be larger than the real input (output) width $p$. The condition $P > p$ induces a padding without any given (measured) values, these values must be assumed for the calculation of the FRT.

## 3. MODIFIED GERCHBERG-SAXTON ALGORITHM

A modified Gerchberg-Saxton algorithm (MGSA) is shown in Fig. 1. The input and output amplitudes $A_1 = \sqrt{I_1}$, $A_2 = \sqrt{I_2}$ are given by an intensity measurement. For the first iteration, the amplitude in the input plane $A_1$ is combined with a random start phase $\phi_1 = \phi_0$ resulting in a complex amplitude $u_1 = A_1 \exp(i\phi_1)$.

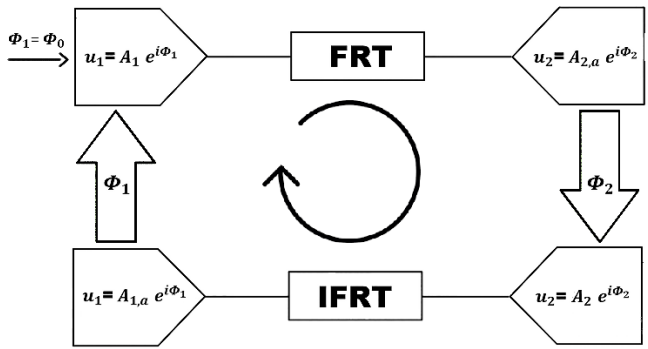

**Fig. 1.** General calculation algorithm of a MGSA in the Fresnel domain.

The corresponding output field $u_2$ at the distance $z$ is calculated with the help of the FRT:

$$u_2 = \frac{1}{i\lambda z} e^{i\frac{\pi r_2^2}{\lambda z}} e^{i2\pi \frac{z}{\lambda}} \mathcal{F}\left\{u_1 e^{ik\frac{\pi r_1^2}{\lambda z}}\right\} = FRT(u_1) \quad (4)$$

Due to the random phase $\phi_0$, the output field $u_2$ is approximated. In the next step the approximated amplitude $A_{2,a} = |u_2|$ is replaced by the real one $A_2$ while keeping the approximated phase:

$$u_2 = A_{2,a} e^{i\phi_2} \rightarrow u_2 = A_2 e^{i\phi_2} \quad (5)$$

Then, the Inverse-Fresnel-Transform (IFRT) will be applied on the modified field $u_2$ to calculate the approximated input field $u_1$:

$$u_1 = i\lambda z\, e^{i\frac{\pi r_1^2}{\lambda z}} e^{-i2\pi \frac{z}{\lambda}} \mathcal{F}^{-1}\left\{u_2 e^{-i\frac{\pi r_2^2}{\lambda z}}\right\} = IFRT(u_2) \quad (6)$$

Analog to $u_2$ the input field $u_1$ is modified by the use of the real input amplitude $A_1$ as:

$$u_1 = A_{1,a} e^{i\phi_1} \rightarrow u_1 = A_1 e^{i\phi_1} \quad (7)$$

The iteration is repeated until the phases $\phi_1$ and $\phi_2$ converge. However, in some cases the MGSA can suffer from non-convergence or stagnation, which is beyond the scope of this work.

## 4. COMPARISON OF SOME METHODS FOR THE USE OF MGSA IN THE FRESNEL DOMAIN

As discussed, for the given physical parameters: $z, \delta x$ and $\lambda$ the width of the computational domain $P$ is fixed and can be larger than the real input and output width $p$. That means a padding without any given amplitude values is required. However, for the numerical calculation some amplitude values must be assumed for the padding points. In Fig. 2 an example for $P > p$, i.e. the width of the computational domain $P$ is larger than the actual width of the image $p$ is shown. In order to show the potential of the methods, two completely independent images are presented. The MGSA will be used to find the phase function for the transfer from the input to the output image and vice versa.

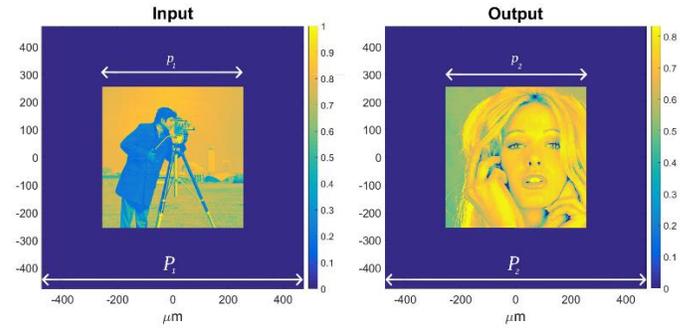

**Fig. 2.** Original input and output image [22] with padding induced by the FRT. The used parameters for all simulations in this paper are $\lambda = 0.633\ \mu m, p = 512\ \mu m, \delta x = 1\ \mu m, z = 1500\ \mu m$.

**Zero-padding**

The common method for solving the aforementioned padding problem is to set the values in the padding zone to zero [23-25]. Fig. 3 shows the reconstructed amplitude and phase according to zero-padding by the use of a MGSA introduced in Fig. 1. Since for the used input, output images and the simulation parameters, after 100 iterations, there was no significant change in the reconstructed amplitudes detected, the termination criterion for all compared methods in this paper was set to 100 iterations in the corresponding MGSA.

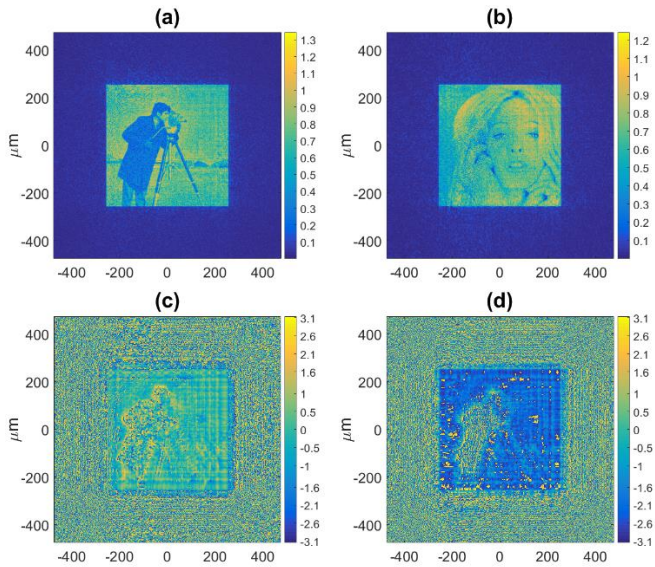

**Fig. 3**. MGSA implementation by Zero–padding, (a) and (b) reconstructed input and output amplitude, (c) and (d) phase functions to transfer the input to the output image and vice versa.

### A. Nonzero-padding

In this approach, the padding zone is filled with a fixed non-zero amplitude value [26,27]. Since the value for the best reconstruction is not known in advance, the MGSA must be repeated in small steps for different amplitude values for an interval $[A_{min}, A_{max}]$. Afterwards the padding value leading to the best reconstruction is selected.

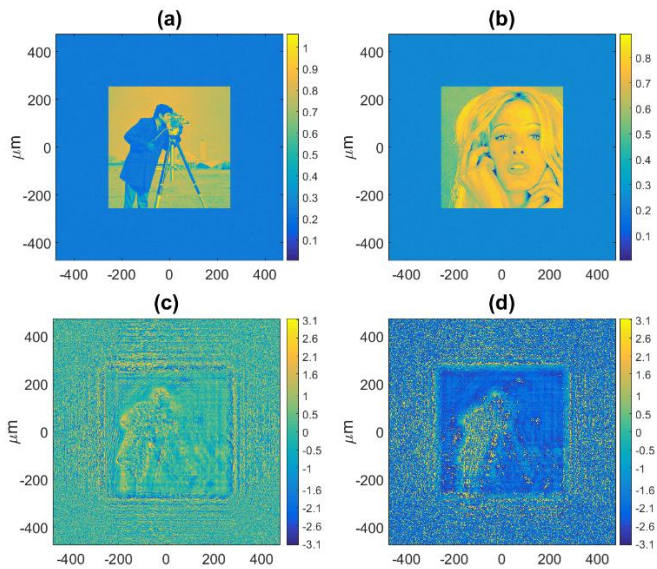

**Fig. 4.** MGSA implementation by nonzero–padding: (a) and (b) reconstructed input and output amplitude respectively. (c) and (d) phase functions to transfer the input to the output image and vice versa.

The drawback of this approach is again, the fixed predefined value for the amplitude in the padding zone. The second disadvantage is a higher numerical effort due to a multiple usage of the modified MGSA. However, compared to the zero-padding it shows better results. Figure 4 demonstrates the reconstructed amplitude and phase for a nonzero padding. The padding value was extracted from the interval [0.1,1] in steps of 0.1 after 10 rounds of the MGSA. After a total number of 1000 iterations (10 x MGSA with 100 iterations), the selected padding amplitude is 0.1. For the nonzero-padding there exists no fundamentally justified criteria for the choice of $A_{min}$, $A_{max}$ and the step interval. Consequently, it might be possible, that the best padding value will not be found by the algorithm.

### B. Variable-padding

Here an approach which allows an almost perfect phase retrieval will be introduced. In principle, the diffraction of the complex amplitude of the input field results in the complex output field. By diffraction the complex input field influences the output plane in the region of the image and in the padding zone. Thus, it is not necessarily restricted to the original size of the input and output image.

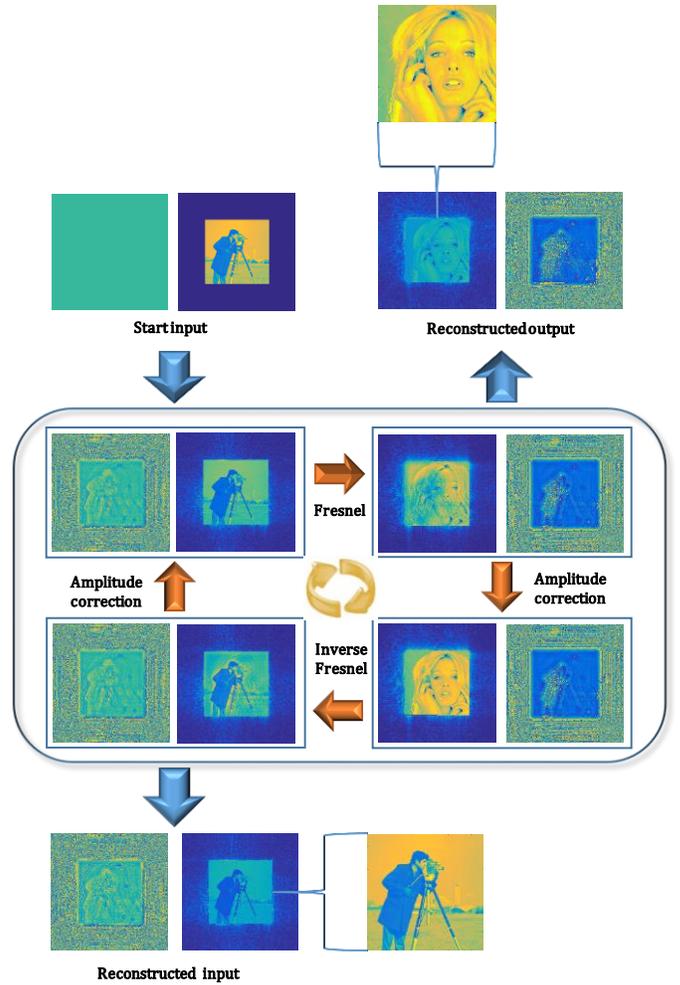

**Fig. 5.** Schematic of the proposed variable-padding MGSA.

Consequently, the amplitude values in the padding zones are usually not zero or constant at all points. Thus, the main idea of the proposed method is the evaluation of the complex amplitude in the entire output computational domain (including the output padding zone) from the complex amplitude in the entire input computational domain (including the input padding zone) and vice versa. Figure 5 demonstrates the proposed algorithm schematically. The start phase is zero $\phi_0(x,y) = 0$ and the start amplitude $u_1$ is the given input amplitude $u_1 = A_1$.

For the calculation of the FRT, the input image has to include a padding zone. Here the amplitude of the pixels in the padding zone will be directly calculated from the diffraction pattern of the respective other plane by the FRT or IFRT and will be modified in each MGSA iteration. For the next MGSA iteration the padding zone will be kept and the area of the image will be replaced by the original image. Therefore, the amplitude values are automatically matched and they are calculated by a physically justified method i.e. the diffraction theory. As for the zero-padding the total number of iterations is just 100.

In Fig. 6 the reconstructed amplitudes and phases, calculated by the implementation of the proposed procedure, are demonstrated. The main differences between original and reconstructed images occur from the fact, that Matlab rescales the colors of the images according to the maximum and minimum values. For the proposed method the maximum values occur in the padding zone.

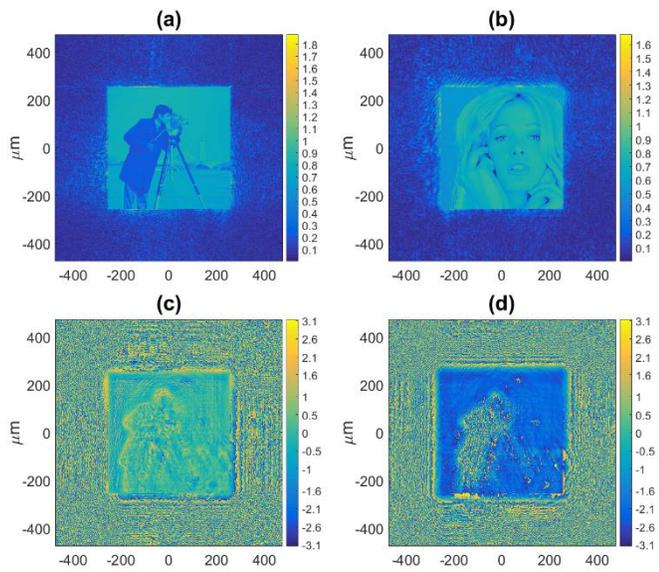

**Fig. 6.** Variable-padding: (a) and (b) reconstructed input and output amplitude. (c) and (d) phase functions to transfer the input to the output image and vice versa.

Figure 7 compares the reconstructed amplitudes of the three methods. For the comparison with the original images the padding zone has been removed.

The images 7(c) and 7(d) are results of zero-padding, whereas 7(e) and 7(f) show the reconstructed amplitudes according to the nonzero-padding. Evidently nonzero-padding exhibit much better reconstruction quality than zero-padding. However, due to the multiple use of the MGSA, the numerical effort was ten times higher. The images 7(g) and 7(h) are the reconstructed input and output amplitudes by the use of the proposed variable-padding. As can be seen, it shows even better results than the nonzero-padding with a computational effort comparable to zero-padding. In order to quantify the reconstruction quality, the correlation coefficients for the three investigated methods have been calculated for the input image and are presented in Fig. 8. The presented correlation gives the similarity between the original input image and the input image calculated from the output image by the MGSA and the founded phase mask. The correlation coefficients depend on the iteration number in the MGSA. For zero- and variable-padding 100 iterations were carried out whereas nonzero-padding has required 10 x 100 iterations. For the plotted graph, only the 100 iterations with the best padding value (0.1) have been shown. As can be seen, in comparison to the two other methods, the proposed algorithm exhibits a better correlation coefficient and has a much faster convergence. For only 10 iterations the correlation coefficient of the variable-padding reached a value of 97.02%, where nonzero-padding with the best value has a coefficient of 89.64% and the correlation value of zero-padding was only around 77.50% after 10 iterations.

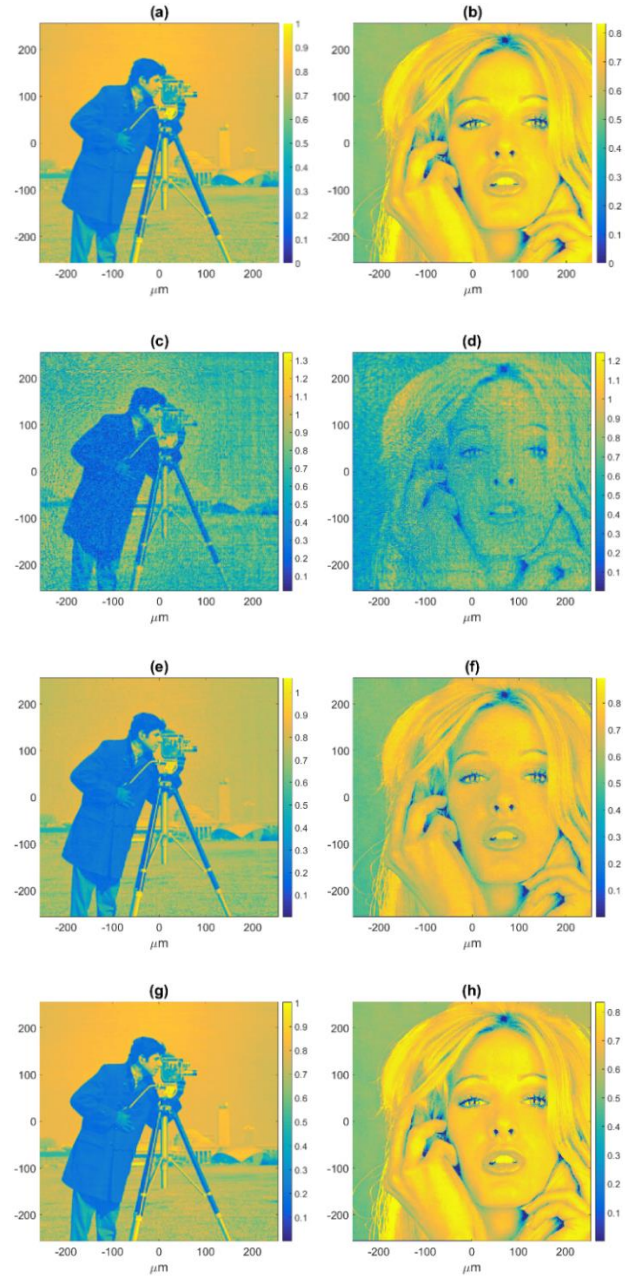

**Fig. 7.** (a) and (b) original amplitude of the input and output. (c), (e) and (g) reconstructed input amplitude by the zero-, nonzero- and variable-padding respectively. (d), (f) and (h) reconstructed output amplitude.

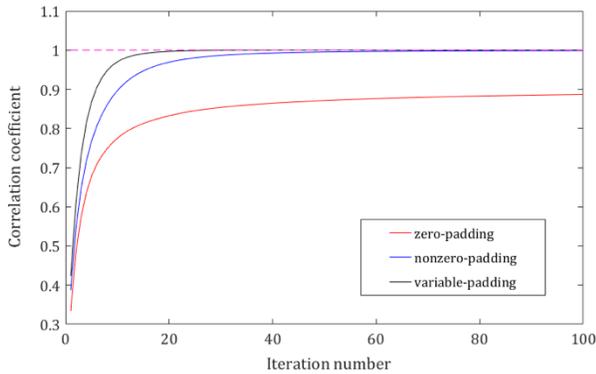

**Fig. 8.** Comparison of the convergence between the original and the computed input image for the three padding methods.

In Tab. 1 the correlation coefficients for the input and output images with its originals, of the three methods are compared after 100 iterations in the MGSA. Here again, for nonzero-padding the MGSA itself was repeated 10 times to find the best padding value. Thus, the total iteration number was 1000.

Table 1. Comparison of the correlation coefficients for the input and output images in %.

|  | Input | Output |
| --- | --- | --- |
| Zero-padding | 88.688806 | 69.67757 |
| Nonzero-padding | 99.877207 | 99.57769 |
| Variable-padding | 99.999992 | 99.999973 |

## 5. CONCLUSION

Here a novel modified Gerchberg-Saxton algorithm in the Fresnel domain was presented and compared to other methods. Compared to zero-padding the method shows a much higher accuracy. Compared to nonzero-padding it shows an accuracy improvement and it is much faster since no iterations for finding the padding values are required. Another advantage of the variable–padding is its much faster convergence. The maximal error for the simulations for the method presented in this paper is only 0.00003 %.

**Acknowledgment**. We acknowledge financial support from "Niedersächsisches Vorab" through "Quantum- and Nano-Metrology (QUANOMET)" initiative within the project NL - 4. We like to thank Thomas Schneider, Stefan Preußler, Ali Dorostkar, Okan Özdemir, Janosch Meier, Arijit Misra and Naghmeh Akbari for their support during the writing of the paper.